\documentclass{appolb}
\usepackage{epsfig}

\begin{document}

\title{Quasi-power law ensembles}
\author{Grzegorz Wilk\thanks{e-mail: wilk@fuw.edu.pl}
\address{National Centre for Nuclear Research,Department of Fundamental Research,\\
 Ho\.za 69, 00-681; Warsaw, Poland}
\and Zbigniew W\l
odarczyk\thanks{e-mail:zbigniew.wlodarczyk@ujk.edu.pl}
\address{Institute of Physics, Jan Kochanowski University,\\
               \'Swi\c{e}tokrzyska 15; 25-406 Kielce, Poland}}

\maketitle

\begin{abstract}
Quasi-power law ensembles are discussed from the perspective of
nonextensive Tsallis distributions characterized by a nonextensive
parameter $q$. A number of possible sources of such distributions
are presented in more detail. It is further demonstrated  that
data suggest that nonextensive parameters deduced from Tsallis
distributions functions $f\left(p_T\right)$, $q_1$, and from
multiplicity distributions (connected with Tsallis entropy),
$q_2$, are not identical and that they are connected via $q_1 +
q_2 = 2$. It is also shown that Tsallis distributions can be
obtained directly from Shannon information entropy, provided some
special constraints are imposed. They are connected with the type
of dynamical processes under consideration (additive or
multiplicative). Finally, it is shown how a Tsallis distribution
can accommodate the log-oscillating behavior apparently seen in
some multiparticle data.
\end{abstract}


\vspace{1cm}

\section{Introduction}
\label{sec:I}

The two most characteristic ensembles is, on one side, the one
resulting in exponential (Boltzmann-Gibbs - (BG)) distributions,
$f_E(X) \sim \exp(-X/T)$, and, on the other side, the one with
power distributions, $f_P(X)\sim X^{-\gamma}$. Both are
encountered in the realm of the high energy multiparticle
production processes investigated in hadronic and nuclear
collisions. They are connected there with, respectively,
nonperturbative {\it soft} dynamics operating at small $X$'s and
described by exponential distributions, $f_E(X)$), and with
perturbative {\it hard} dynamics, responsible for large $X$'s and
described by power distributions, $f_P(X)$. These two types of
dynamics are investigated separately. It is usually assumed that
they operate in distinct parts of phase space of  $X$, separated
by $X=X_0$. However, it was found recently that the new high
energy data covering the whole of phase space (cf., for example,
\cite{CMS,ALICE,ATLAS}) are best fitted by a simple, quasi-power
law formula extrapolating between both ensembles \cite{CM,UA1,H}:
\begin{eqnarray}
  H(X) = C\cdot \left( 1 + \frac{X}{nX_0}\right)^{-n}
  \longrightarrow
  \left\{
 \begin{array}{l}
  \exp\left(-\frac{X}{X_0}\right)\quad \, \, \, {\rm for}\ X \to 0, \smallskip\\
  X^{-n}\qquad \qquad{\rm for}\ X \to \infty,
 \end{array}
 \right .
 \label{eq:H}
\end{eqnarray}
This formula coincides with the so called Tsallis nonextensive
distribution \cite{Tsallis} for $n=1/(q-1)$,
\begin{equation}
h_q(X) = C_q\left[ \!1-\! (1-q)\frac{X}{X_0}
\right]^{\frac{1}{1-q}} \stackrel{def}{=}
C_q\exp_q\left(-\frac{X}{X_0}\right)
 \stackrel{q \rightarrow 1}{\Longrightarrow} C_1\exp
\left(-\frac{X}{X_0}\right).\label{eq:Tsallis}
\end{equation}
This is the distribution we shall shall concentrate on and
discuss. In Section \ref{sec:Examples} we shall discuss some
examples of processes leading to such distributions. It depends on
the nonextensivity parameter $q$ and this can be different
depending on whether it arises from a Tsallis distribution ($q_1$)
or from the nonextensive Tsallis entropy ($q_2$). Both are
connected by $q_1 +q_2 =2$ and this relation seems to be confirmed
experimentally. This is presented in Section \ref{sec:duality}. In
Section \ref{sec:Shannon} we shall discuss necessary conditions
for obtaining a Tsallis distribution from Shannon information
entropy. Section \ref{sec:logosc} demonstrates that a Tsallis
distribution can also accommodate the log-periodic oscillations
apparently observed in high energy data. Our conclusions and a
summary are presented in Section \ref{sec:Summary}.

\section{Some examples of mechanisms leading to Tsallis
distributions} \label{sec:Examples}

In many practical applications, a Tsallis distribution is derived
from Tsallis statistics based on his nonextensive
entropy\footnote{Cf. \cite{Sq} for most recent work with
references; thermodynamical consistency of such an approach can be
found in \cite{consistency}. We shall not discuss this point
here.},
\begin{equation}
S_q = -\sum p_i \ln_q p_i\quad{\rm where}\quad \ln_q x = \frac{
p^{q-1} -1}{q-1}\label{eq:Sqlnq}
\end{equation}
On the other hand, there are even more numerous examples of
physical situations not based on $S_q$ and still leading to
quasi-power distributions in the Tsallis form. In what follows, we
shall present some examples of such mechanisms, concentrating on
those which allow for an interpretation of the parameter $q$.

\subsection{Superstatistics}
\label{sec:Ss}

The first example is {\it superstatistics} \cite{SuperS} (cf, also
\cite{WW,BJ}) based on the property that a gamma-like fluctuation
of the scale parameter in exponential distribution results in the
$q$-exponential Tsallis distribution with $q>1$ (cf. Eq.
(\ref{eq:Tsallis})). The parameter $q$ defines the strength of
such fluctuations, $q = 1 + Var(X)/<X>^2$. From the thermal
perspective, it corresponds to a situation in which the heat bath
is not homogeneous, but has different temperatures in different
parts, which are fluctuating around some mean temperature $T_0$.
It must be therefore described by two parameters: a mean
temperature $T_0$ and the mean strength of fluctuations, given by
$q$. As shown in \cite{WWrev}, this allows for further
generalization to cases where one also has an energy transfer
to/from heat bath. The scale $T$ in the Tsallis distribution
becomes then $q$-dependent:
\begin{equation}
T = T_{eff} = T_0 + (q-1)T_V \label{eq:Teff}
\end{equation}
Here the parameter $T_V$ depends on the type of energy transfer,
cf. \cite{WWTout,WWTin} for illustrative examples from,
respectively, nuclear collisions and cosmic ray physics.

\subsection{Preferential attachment}
\label{sec:Sn}

The second example is the {\it preferential attachment approach}
(used in stochastic networks \cite{nets}). Here the system under
consideration exhibits correlations of the preferential attachment
type (like, for example, "rich-get-richer" phenomenon in networks)
and the scale parameter depends on the variable under
consideration. If $x_0\rightarrow x_0'(x)=x_0+(q-1)x$ then the
probability distribution function, $f(x)$, is given by an equation
the solution of which is a Tsallis distribution (again, with
$q>1$):
\begin{equation}
\frac{df(x)}{dx} = -
\frac{1}{x_0'(x)}f(x)~~~\Longrightarrow~~~f(x) =
\frac{2-q}{x_0}\left[ 1 -
(1-q)\frac{x}{x_0}\right]^{\frac{1}{1-q}}. \label{eq:nets}
\end{equation}
For $x_0'(x) = x_0$ one again gets the usual exponential
distribution.

\subsection{Multiplicative noise}
\label{sec:Mn}

Consider now a {\it Tsallis distribution from multiplicative
noise} \cite{WW,BJ}. We start from the Langevin equation
\cite{BJ},
\begin{equation}
\frac{dp}{dt} + \gamma(t)p = \xi(t) \label{eq:Le}
\end{equation}
where $\gamma(t)$ and $\xi(t)$ denote stochastic processes
corresponding to, respectively, multiplicative and additive
noises. This results in the following Fokker-Planck equation for
the distribution function $f$,
\begin{equation}
\frac{\partial f}{\partial t} = - \frac{\partial \left( K_1
f\right)}{\partial p} + \frac{\partial^2 \left( K_2
f\right)}{\partial p^2}. \label{eq:FPe}
\end{equation}
Stationary $f$ satisfies
\begin{equation}
\frac{d \left( K_2 f\right)}{dp} = K_1 f, \label{eq:K1vK2}
\end{equation}
with
\begin{equation}
K_1 = \langle \xi\rangle - \langle \gamma\rangle p\quad {\rm
and}\quad K_2 = Var(\xi) - 2 Cov(\xi, \gamma) p + Var(\gamma )
p^2. \label{eq:K1K2}
\end{equation}
In the case of no correlation between noises and no drift term due
to additive noise (i.e., for $Cov(\xi,\gamma) = \langle \xi\rangle
= 0$ \cite{AT}) its solution is a Tsallis distribution for $p^2$,
\begin{equation}
f(p) = \left[1 + (q - 1)\frac{p^2}{T}\right]^{\frac{q}{1-q}}~~{\rm
with}~~ T = \frac{2Var(\xi)}{\langle \xi\rangle};~~q = 1 +
\frac{2Var(\gamma)}{\langle \gamma\rangle}. \label{eq:solutionpp}
\end{equation}
However, if we insist on a solution in the form of
\begin{equation}
f(p) = \left[1 + \frac{p}{nT}\right]^n,~~n = \frac{1}{q-1},
\label{eq:singlep}
\end{equation}
Eq. (\ref{eq:K1vK2}) has to be replaced by
\begin{equation}
K_2(p) = \frac{nT+p}{n}\left[ K_1(p) - \frac{dK_1(p)}{dp}\right].
\label{eq:K1vK2p}
\end{equation}
One then gets $f(p)$ in the form of a Tsallis distribution,
(\ref{eq:singlep}), but with
\begin{equation}
n = 2 + \frac{\langle \gamma\rangle}{Var(\gamma)}\quad{\rm
or}\quad q = 1 + \frac{Var(\gamma)}{\langle \gamma\rangle + 2
Var(\gamma)} \label{eq:n-q}
\end{equation}
and  with $q$-dependent $T$ (reminiscent of $T_{eff}$ from Eq.
(\ref{eq:Teff}) discussed before, cf. \cite{WW_AIP}):
\begin{equation}
T(q) = (2-q)\left[ T_0 + (q-1)T_1\right]\quad{\rm with}~~
T_0=\frac{Cov(\xi,\gamma)}{\langle \gamma\rangle},~~T_1 =
\frac{\langle \xi\rangle}{2\langle \gamma\rangle}. \label{eq:Teff}
\end{equation}

\subsection{All variables fixed}
\label{sec:AllF}

Let us now remember that the usual situation in statistical
physics is that out of three variables considered, energy $U$,
multiplicity $N$ and temperature $T$, two are fixed and one
fluctuates. Fluctuations are then given by gamma distributions
\cite{WWcov} (in the case of multiplicity distributions where $N$
are integers, they become Poisson distributions) and only in the
thermodynamic limit ($N \rightarrow \infty$) does one get them in
the form of Gaussian distributions, usually discussed in
textbooks. In \cite{WWcov} we discussed in detail situations when
two or all three variables fluctuate. If all are fixed we have a
distribution of the type of
\begin{equation}
f(E) = \left(1 - E/U\right)^{N-2}. \label{eq:ConstN}
\end{equation}
This is nothing else but a Tsallis distribution with $q
=(N-3)/(N-2) < 1$~~~\footnote{Actually, such distributions emerges
directly from calculus of probability for situation known as {\it
induced partition} \cite{IndPart}. In short: $N-1$ randomly chosen
independent points $\left\{ U_{1},\dots, U_{N-1}\right\}$ breaks
segment $(0,U)$ into $N$ parts, length of which is distributed
according to Eq. (\ref{eq:ConstN}). The length of the $k^{th}$
such part corresponds to the value of energy $E_k = U_{k+1}-U_k$
(for ordered $U_k$). One could think of some analogy in physics to
the case of random breaks of string in $N-1$ points in the energy
space. Notice that induced partition differs from {\it successive
sampling} from the uniform distribution, $E_k\in \left[
0,U-E_1-E_2- \dots  -E_{k-1} \right]$, which results in $f(E)
=1/E$~~\cite{ONEoverE}. }.

\subsection{Conditional probability}
\label{sec:Cp}

For the {\it constrained systems} one gets $q < 1$. For example,
if we have $n$ independent energies,
$\left\{E_{i=1,\dots,N}\right\}$, then each of them is distributed
according to the Boltzman distribution, $g_i\left(E_i\right) =
(1/\lambda)\exp\left(-E_i/\lambda\right)$ (and their sum,
$E=\sum_{i=1}^{N}E_i$, is distributed according to gamma
distribution, $g_N(E) = 1/[\lambda(N-1)](E/\lambda)^{N-1}
\exp(-E/\lambda)$). However, if the available energy is limited,
$E=N\alpha= const$, then the resulting {\it conditional
probability} becomes a Tsallis distribution with $q <1$
\footnote{One could get a Tsallis-like distribution with $q
>1$ only if the scale parameter $\lambda$ would fluctuate in the
same way as in in the case of superstatistics, see \cite{WWreV}.}:
\begin{eqnarray}
f\left( E_i|E=N\alpha \right) &=& \frac{g_1\left( E_i \right)
g_{N-1}\left( N\alpha-E_i \right )}{g_N (N\alpha)} =
\frac{(N-1)}{N\alpha}\left(1 -
\frac{1}{N}\frac{E_i}{\alpha}\right)^{N-2} = \nonumber\\
&=& \frac{2-q}{\lambda}\left[ 1 - (1 - q) \frac{E_i}{\lambda}
\right]^{\frac{1}{1 - q}},
\label{eq:constraints}\\
&& q = \frac{N-3}
{N-2} < 1, \qquad \qquad \lambda = \frac{\alpha
N}{N-1}. \label{eq:c1}
\end{eqnarray}

\subsection{Statistical physics}
\label{sec:SPc}

We end this part by a reminder of how Tsallis distribution with
$q<1$ arises from {\it statistical physics considerations}.
Consider an isolated system with energy $U=const$ and with $\nu$
degrees of freedom ($n$ particles). Choose single degree of
freedom with energy $E$ (i.e., the remaining, or reservoir, energy
is $E_r = U - E$). If this degree of freedom is in a single, well
defined, state then the number of states of the whole system is
$\Omega(U-E)$ and probability that the energy of the chosen degree
of freedom is $E$ is $P(E) \propto \Omega(U-E)$. Expanding (slowly
varying) $\ln \Omega(E)$ around $U$,
\begin{equation}
\ln \Omega(U-E) = \sum_{k=0}^{\infty} \frac{1}{k!}
\frac{\partial^{(k)} \ln \Omega}{\partial E_r^{(k)}}, \quad {\rm
with}\quad \beta = \frac{1}{k_B T} \stackrel{def}{=}
\frac{\partial \ln \Omega\left(E_r\right)}{\partial
E_r},\label{eq:deriv}
\end{equation}
and (because $E<<U$) keeping only the two first terms one gets
\begin{equation}
\ln P(E) \propto \ln \Omega(E) \propto - \beta E,\quad {\rm or}
\quad P(E) \propto \exp( - \beta E), \label{eq:BoltzmannD}
\end{equation}
i.e., a Boltzmann distribution (or $q=1$). On the other hand,
because one usually expects that $\Omega\left(E_r\right) \propto
\left(E_r/\nu\right)^{\alpha_1 \nu - \alpha_2}$ (where
$\alpha_{1,2}$ are of the order of unity and we put $\alpha_1 = 1$
and, to account for diminishing the number of states in the
reservoir by one, $\alpha_2 = 2$) \cite{Reif}, one can write
\begin{equation}
\frac{\partial^k \beta}{\partial E_r^k} \propto (-1)^k k!
\frac{\nu - 2}{E^{k+1}_r} = (-1)^k k! \frac{\beta^{k-1}}{(\nu -
2)^k} \label{eq:FR}
\end{equation}
and write the full series for probability of choosing energy $E$:
\begin{eqnarray}
P(E)&\propto& \frac{\Omega(U-E}{\Omega(U)} = \exp\left[
\sum_{k=0}^{\infty}\frac{(-1)^k}{k+1}\frac{1}{(\nu - 2)^k}(- \beta
E)^{k+1}\right]=\nonumber\\
&=& C\left(1 - \frac{1}{\nu - 2}\beta E\right)^{(\nu - 2)}
=\nonumber\\
&=& \beta(2-q)[1 - (1-q)\beta E]^{\frac{1}{1-q}};
\label{eq:statres}
\end{eqnarray}
(where we have used the equality $\ln(1+x) =
\sum_{k=0}^{\infty}(-1)^k[x^{k+1}/(k+1)]$). This result, with $q=
1 - 1/(\nu -2) \leq 1$, coincides with results from conditional
probability and the induced partition.

\subsection{Fluctuations of multiplicity $N$}
\label{sec:FN}

Constant values of $U$, $N$ and $T$ result in Eq.
(\ref{eq:ConstN}) and $q < 1$. To get larger values of $q$, one
has to allow for fluctuations of one of the variables: $U$, $N$ or
$T$. In superstatistics \cite{SuperS,WW,BJ} it was $T$ that was
fluctuaing, let us now consider the example of fluctuating $N$. It
means that, whereas for fixed $N$ (to simplify notation we changed
$N-2$ to $N$ here)
\begin{equation}
f_N(E) = \left(1 - \frac{E}{U}\right)^N\qquad {\rm and}\qquad U =
\sum E = const, \label{eq:fixedN}
\end{equation}
for $N$ fluctuating according to some $P(N)$ the resulting
distribution is
\begin{equation}
f(E) = \sum f_N(E) P(N). \label{eq:resdistr}
\end{equation}
The most characteristic for our purposes are situations provided
by the, respectively, {\it Binomial Distribution} (BD),
Eq.(\ref{eq:BD}), {\it Poissonian Distribution} (PD),
Eq.(\ref{eq:PD}) and by the {\it Negative Binomial Distributions}
(NBD), Eq.(\ref{eq:NBD}) (cf. \cite{WWjets}):
\begin{eqnarray}
P_{BD}(N) &=& \frac{M!}{N!(M-N)!}
\left(\frac{<N>}{M}\right)^N\left(1 -
\frac{<N>}{M}\right)^{M-N}; \label{eq:BD}\\
P_{PD}(N)&=& \frac{<N>^N}{N!} e^{-\langle N\rangle} \label{eq:PD}\\
P_{NBD}(N) &=& \frac{\Gamma(N+k)}{\Gamma(N+1)\Gamma(k)}\left(
\frac{<N>}{k}\right)^N \left( 1 + \frac{<N>}{k}\right)^{-k-N}.
\label{eq:NBD}
\end{eqnarray}
They lead, respectively, to Tsallis distribution with $q=1-1/M
<1$, Eq.(\ref{eq:BDf}), to exponential Boltzmann distribution,
Eq.(\ref{eq:PDf}) with $q=1$, and to Tsallis distribution with $q
= 1 + 1/k
>1$, Eq.(\ref{eq:NBDf}) ($\beta = \langle N\rangle/U$):
\begin{eqnarray}
f_{BD}(E) &=& \left(1 - \frac{\beta
E}{M}\right)^M,  \label{eq:BDf}\\
f_{PD}(E) &=& \exp( - \beta E), \label{eq:PDf}\\
f_{NBD}(E) &=& \left( 1 + \frac{\beta E}{k}\right)^{-k}.
\label{eq:NBDf}
\end{eqnarray}
Note that in all three cases
\begin{equation}
q - 1 = \frac{Var(N)}{<N>^2} - \frac{1}{<N>}. \label{eq:FluctN}
\end{equation}
It is natural that for BD where $Var(N)/<N> < 1$ one has $q <1$,
for PD with $Var(N)/<N> =1$ also $q=1$ and for NBD where
$Var(N)/<N> > 1$ one has $q>1$ (cf., also case $U=const$ and
$T=const$ considered in \cite{WWcov}).

Fluctuations of $N$ can be translated into fluctuations of $T$.
Notice first that \cite{WWrev} (cf. also \cite{NBDder}) NBD, $P(N)
=\Gamma(N+k)/[\Gamma(N+1)\Gamma(k)]\cdot\gamma^k(1+\gamma)^{-k-N}$,
arises also if in the Poisson multiplicity distribution, $P(N)=
\bar{N}^N e^{-{\bar{N}}}/N!$, one fluctuates the mean multiplicity
$\bar{N}$ using gamma distribution $f(\bar{N}) =
\gamma^k{\bar{N}}^{k-1}/\Gamma(k)\cdot e^{-\gamma \bar{N}}$ with
$\gamma = k/<\bar{N}>$~~\footnote{We have two type of averages
here: $\bar{X}$ means average value in a given event whereas $<X>$
denotes averages over events (or ensembles).}. Now, identifying
fluctuations of mean $\bar{N}$ with fluctuations of $T$, one can
express the above observation via fluctuations of temperature.
Noticing that $\bar{\beta} = \bar{N}/U$ (i.e., that $<\bar{N}> =
U<\bar{\beta}>$ and $\gamma = k/[U<\bar{\beta}>]$) one can rewrite
$f(\bar{N})$ as
\begin{eqnarray}
f\left( \bar{\beta}\right) &=&
\frac{k}{<\bar{\beta}>\Gamma(k)}\left(
\frac{k\bar{\beta}}{<\bar{\beta}>}\right)^{k-1}
\exp\left(-\frac{k\bar{\beta}}{<\bar{\beta}>}\right)=\nonumber\\
&=& \frac{ \left(\frac{1}{q-1}\frac{\bar{\beta}}{{<\bar{\beta}>}}
\right)^{\frac{1}{q - 1} -
1}}{(q-1)<\bar{\beta}>\Gamma\left(\frac{1}{q-1}\right)}
\exp\left(-\frac{1}{q-1}\frac{\bar{\beta}}{<\bar{\beta}>}\right).
\label{eq:NtoT}
\end{eqnarray}
And this is just a gamma distribution describing fluctuations of
$\beta = 1/T$ discussed in \cite{WW}.

There is more to the physical meaning of $q$. Because $U = <N>T$
the heat capacity $C$ can be written as $1/C= dU/dT = <N>$.
However, because in our case $U=const$, i.e., $Var(N)/\langle
N\rangle^2 = Var(\beta)/\langle \beta\rangle^2$ ~(or $<N> \sim
<\beta>$ and $Var(N)\sim Var(\beta)$), the formula
$q-1=Var(N)/<N>^2 - 1/<N>$ obtained before coincides with a
similar formula obtained in \cite{C_Biro}:
\begin{equation}
q - 1 = \frac{Var(\beta)}{<\beta>^2} - \frac{1}{C} \simeq
\frac{Var(T)}{<T>^2} - \frac{1}{C}.\label{eq:C_Biro}
\end{equation}
This can be confronted with $q-1=Var(\bar{N})/<\bar{N}>^2$ from
\cite{WWrev}, which, because of the form of $f(\bar{\beta})$ in
Eq. (\ref{eq:NtoT}), can be rewritten approximately as
\begin{equation}
q - 1 = \frac{Var(\bar{\beta})}{<\bar{\beta}>^2} \simeq
\frac{Var(\bar{T})}{{<\bar{T}>}^2}.
\end{equation}
\begin{figure} [h]
\begin{center}
\includegraphics[scale=0.38]{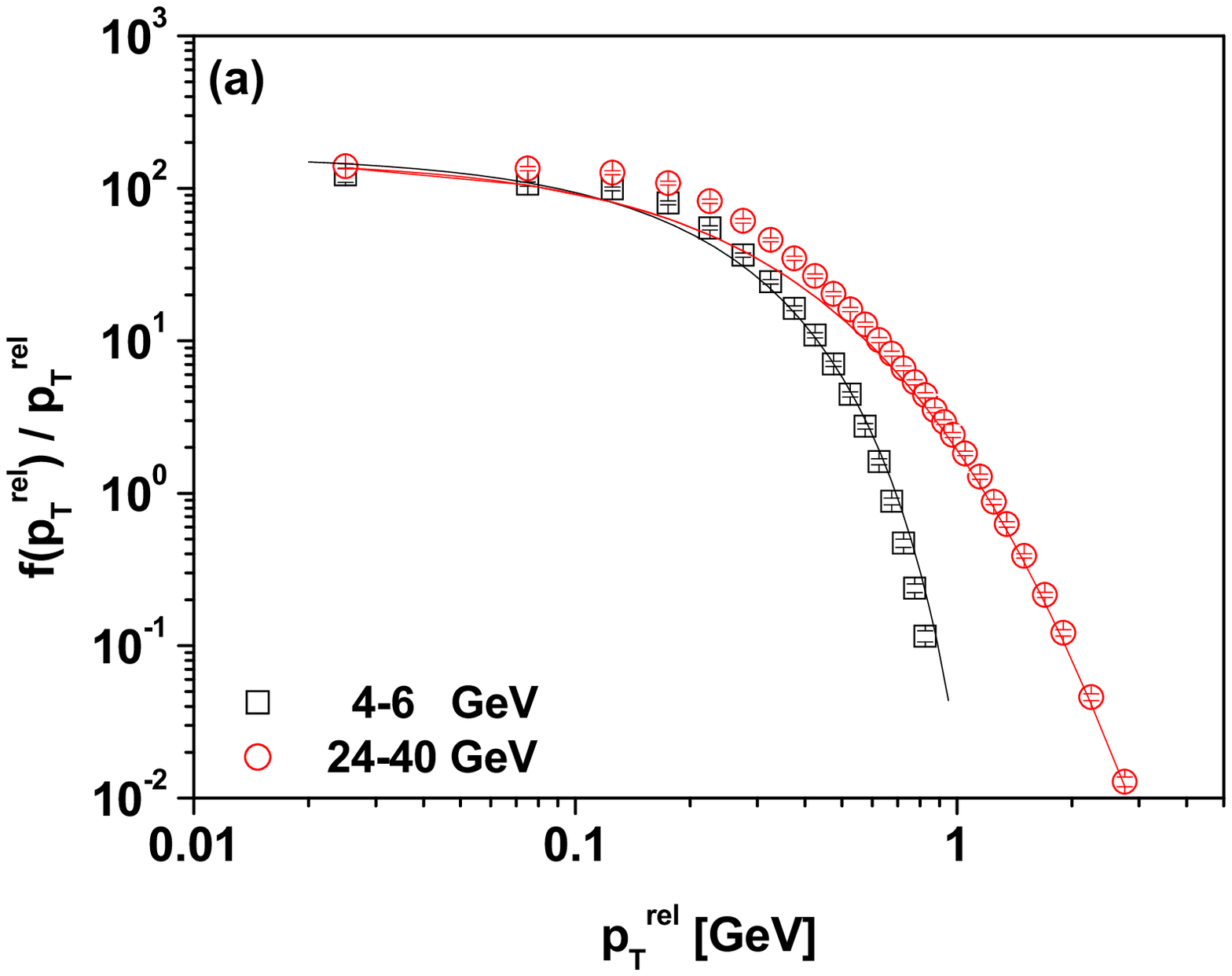}
\includegraphics[scale=0.38]{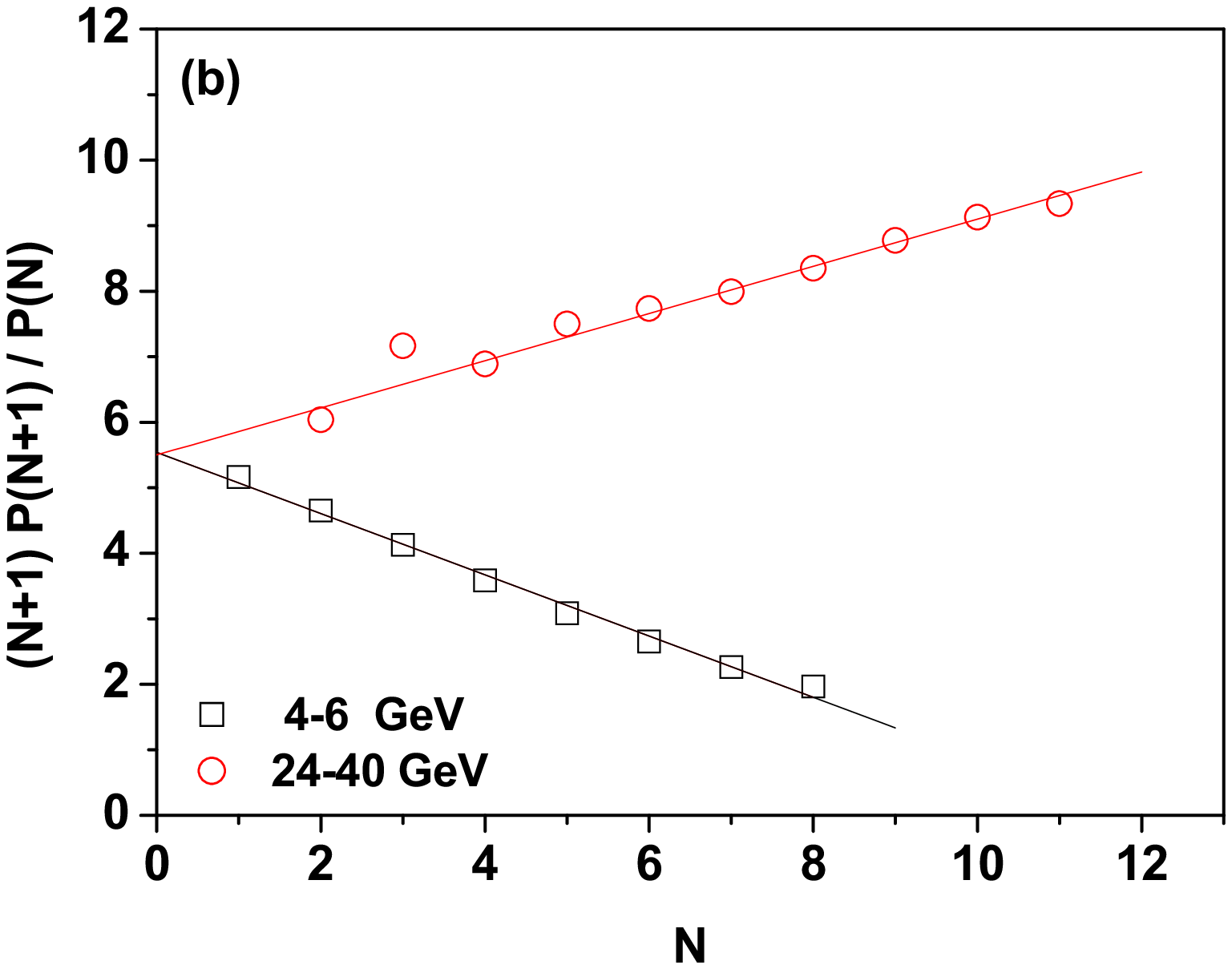}
\end{center}
\caption{ (Color online)  Examples of $q < 1$ or $q > 1$. $(a)$
Distributions of $ p^{rel}_T = \left | \vec{p}\times
\vec{p_{jet}}\right |/
 \left | \vec{p}_{jet}\right |$  for  particles inside the jets
with different values of $\vec{ p}_{jet}$. Distributions are
fitted using a Tsallis distribution (\ref{eq:Tsallis}) with $T =
0.18$.  $(b)$ $(N+1) P(N+1)/P(N) = a + b N $ as function of
multiplicity $N$ in jets with different values of $\vec{p}_{jet}$
as presented in $(a)$. Depending on the range of phase space
covered, the corresponding values of $q$ are $0.885$ for
$\vec{p}_{jet} \in (4-6)$ Gev and $1.094$ for $\vec{p}_{jet} \in
(24-40)$ GeV. Data are from \cite{ATLAS,ATLAS1} and from the
Durham HepData Project;
http://hepdata.cedar.ac.uk/view/irn9136932.
 } \label{Examples}
\end{figure}

\section{Tsallis distribution from Shannon entropy}
\label{sec:Shannon}

Tsallis distribution is usually derived either from Tsallis
entropy (via the MaxEnt variational approach, not discussed here)
or from some dynamical considerations, some examples of which are
presented in this paper. However, it turns out that it also
emerges in a quite natural (in the same MaxEnt approach) way from
Shannon entropy, provided one imposes the right constraints. In
fact, as shown in \cite{TfromS}, one can establish a
transformation between these two variational problems and shown
that they contain the same information. This means that the two
approaches seem to be equivalent, one can either use Tsallis
entropy with relatively simple constraints, or the Shannon entropy
with rather complicated ones (cf., for example, a list of possible
distributions one can get in this way \cite{GR}).

In general, Shannon entropy for some probability density $f(x)$,
$S = - \int dx f(x)\ln[f(x)]$, supplied with constraint $<h(x)> =
\int dx f(x)h(x) = const$, where $h(x)$ is some function of $x$,
subjected to the usual MaxEnt variational procedure, results in
the following form of $f(x)$:
\begin{equation}
f(x) = \exp\left[ \lambda_0 + \lambda h(x)\right],
\label{eq:TfromS}
\end{equation}
with constants $\lambda_0$ and $\lambda$ calculated from the
normalization of $f(x)$ and from the constraint equation. It is
now straightforward to check that
\begin{equation}
<z> = z_0 = \frac{q-1}{2-q}\qquad {\rm  where}\qquad  z =
\ln\left[1 - (1 - q)\frac{E}{T_0}\right] \label{eq:constrT}
\end{equation}
results in $f(z) = (1/z_0)\exp\left( -z/z_0\right)$ which
translates to (remembering that $f(z)dz=f(E)dE$) a Tsallis
distribution
\begin{equation}
\! f(E) = \frac{1}{\left(1\! +\! z_0\right)T_0}\left( 1 +
\frac{z_0}{1+z_0}\frac{E}{T_0}\right)^{ - \frac{1 + z_0}{z_0}} =
\frac{2\! -\! q}{T_0}\left[ 1 - (1 -
q)\frac{E}{T_0}\right]^{\frac{1}{1 - q}}. \label{eq:TfS}
\end{equation}
The parameter $T_0$ can be deduced from the additional condition
which must be imposed, namely from the assumed knowledge of the
$\langle E\rangle$ (notice that in the case of BG distribution
this would be the only condition).

So far the physical significance of the constraint
(\ref{eq:constrT}) is not fully understood. Its form can be
deduced from the idea of varying scale parameter in the form of
the preferential attachment, Eq. (\ref{eq:nets}), which in present
notation means $T \rightarrow T(E)=T_0+(q-1)E$. As shown in
(\ref{eq:nets}) it results in Tsallis distribution (\ref{eq:TfS}).
This suggest the use of $z =\ln\left[ T(E)/T_0\right]$ constrained
as in Eq. (\ref{eq:constrT}). In such approach $\ln f(E) = -
[1/(q-1)]\ln [T(E)] + [(2-q)/(q-1)]\ln \left( T_0\right)$ and,
because $S = - \langle \ln f(E)\rangle$, therefore $S = 1/(2-q) +
\ln \left( T_0\right)$ for Tsallis distribution becoming $S = 1 +
\ln \left( T_0\right)$ for Boltzmann-Gibbs (BG) distribution
($q=1$).

It is interesting that the constraint (\ref{eq:constrT}) seems to
be natural a for multiplicative noise described by the Langevine
equation: $dp/dt + \gamma(t)p=\xi(t)$, with traditional
multiplicative noise $\gamma (t)$ and additive noise (stochastic
processes) $\xi(t)$) (see \cite{WW_AIP} for details). In fact,
there is a connection between the kind of noise in this process
and the condition imposed in the MaxEnt approach. For processes
described by an  additive noise, $dx/dt = \xi(t)$, the natural
condition is that imposed on the arithmetic mean, $<x>= c+\langle
\xi\rangle t$, and it results in the exponential distributions.
For the multiplicative noise, $dx/dt = x\gamma(t)$, the natural
condition is that imposed on the geometric mean, $ <\ln x> =
c+\langle \gamma\rangle t$, which results in a power law
distribution \cite{R}.

\section{Tsallis entropy vs Tsallis distributions ($q_1 + q_2 = 2$)}
\label{sec:duality}

One has to start with some explanatory remarks. The Tsallis
distribution can be also obtained via MaxEnt procedure from
Tsallis entropy,
\begin{equation}
S_q = - \frac{1}{1-q}\sum\left( 1 - p_i^q\right) = - \langle \ln_q
p_i\rangle_q = - \langle \ln_{2-q} p_i\rangle_{q=1}, \label{eq:Sq}
\end{equation}
where $\ln_q x =\left(x^{1-q}-1\right)/(1-q)$ and $\langle
x\rangle_q = \sum p_i^q x_i$. Now, depending on the condition
imposed one gets from $S_q$
\begin{eqnarray}
{\rm either}\quad f(x) &=& q [1+(1-q)x]^{-\frac{1}{1-q}}\quad {\rm
for}\quad\langle x\rangle_1 \label{eq:cond1}\\
{\rm or}\quad f(x) &=& (2-q)[1+(q-1)x]^\frac{1}{1-q}\quad {\rm
for}\quad\langle x\rangle_q. \label{eq:condq}
\end{eqnarray}
However, after replacement of $q$ by $q_1 = 2-q$, the distribution
(\ref{eq:cond1}) becomes the usual Tsallis distribution
(\ref{eq:condq}). Therefore one encounters an apparent puzzle,
namely the $q_1$ of Tsallis distribution does not coincides with
the $q_2$ of corresponding Tsallis entropy, instead they are
connected by relation $q_1 + q_2 = 2$. The natural question
therefore arises: is such a relation seen in data? As shown in
\cite{WWCyprus} that seems really be the case, at least
quantitatively. This is seen when comparing $q = q_1$ obtained
from data on $p_T$ {\it distributions} (cf. Fig.
\ref{Duality}$(a)$) to $q=q_2$ obtained from data on {\it
multiplicities} in $p-A$ collisions assuming that entropy is
proportional to the number of particles produced (cf. Fig.
\ref{Duality} $(b)$). Whereas $q_1$ is deduced from a Tsallis
distribution taken in one of the forms discussed above, $q_2$ is
deduced directly from the corresponding entropy $S_q$ of the $p-A$
collision.  Assume that such collision can be adequately described
by a superposition model in which the main ingredients are $\nu$
nucleons which have interacted at least once \cite{Superposition}.
Assume further that they are identical and independent and produce
$n_i$ secondaries of each other. As a result a
$N=\sum_{i=1}^{\nu}n_i$ are produced in one collision and the mean
multiplicity is $\langle N\rangle = \langle \nu\rangle \langle
n_i\rangle$ (where $\langle \nu\rangle$ is the mean number of
nucleons participating in the collision and $\langle n_i\rangle$
the mean multiplicity in $i^{th}$ elementary collision. The
corresponding entropy $S^{(\nu)}_q$ of such process will then be
$q$-sum of $\nu$ entropies $S^{(1)}_q$ of individual collisions
and is given by:
\begin{figure} [t]
\begin{center}
\includegraphics[scale=0.35]{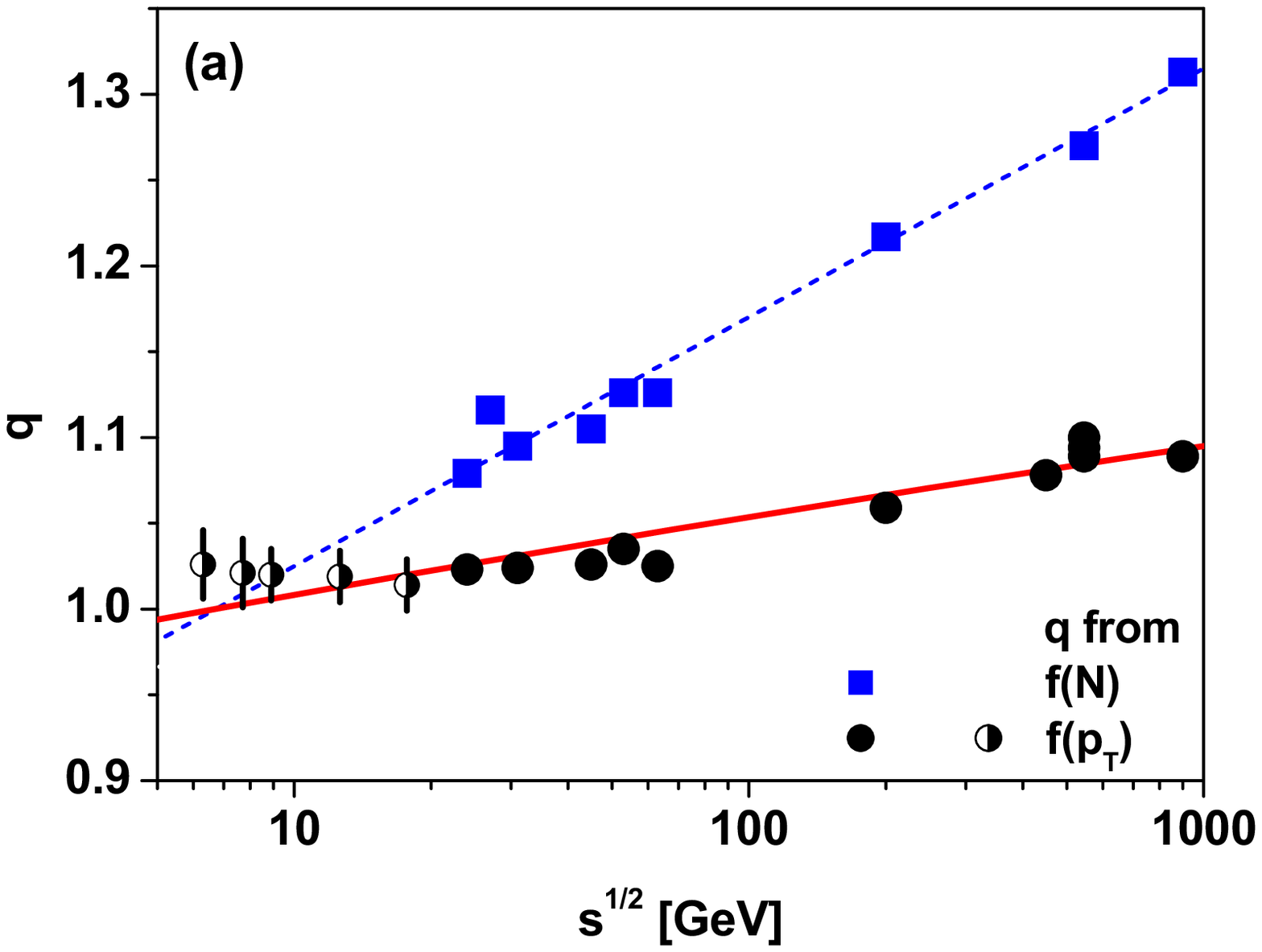}
\includegraphics[scale=0.35]{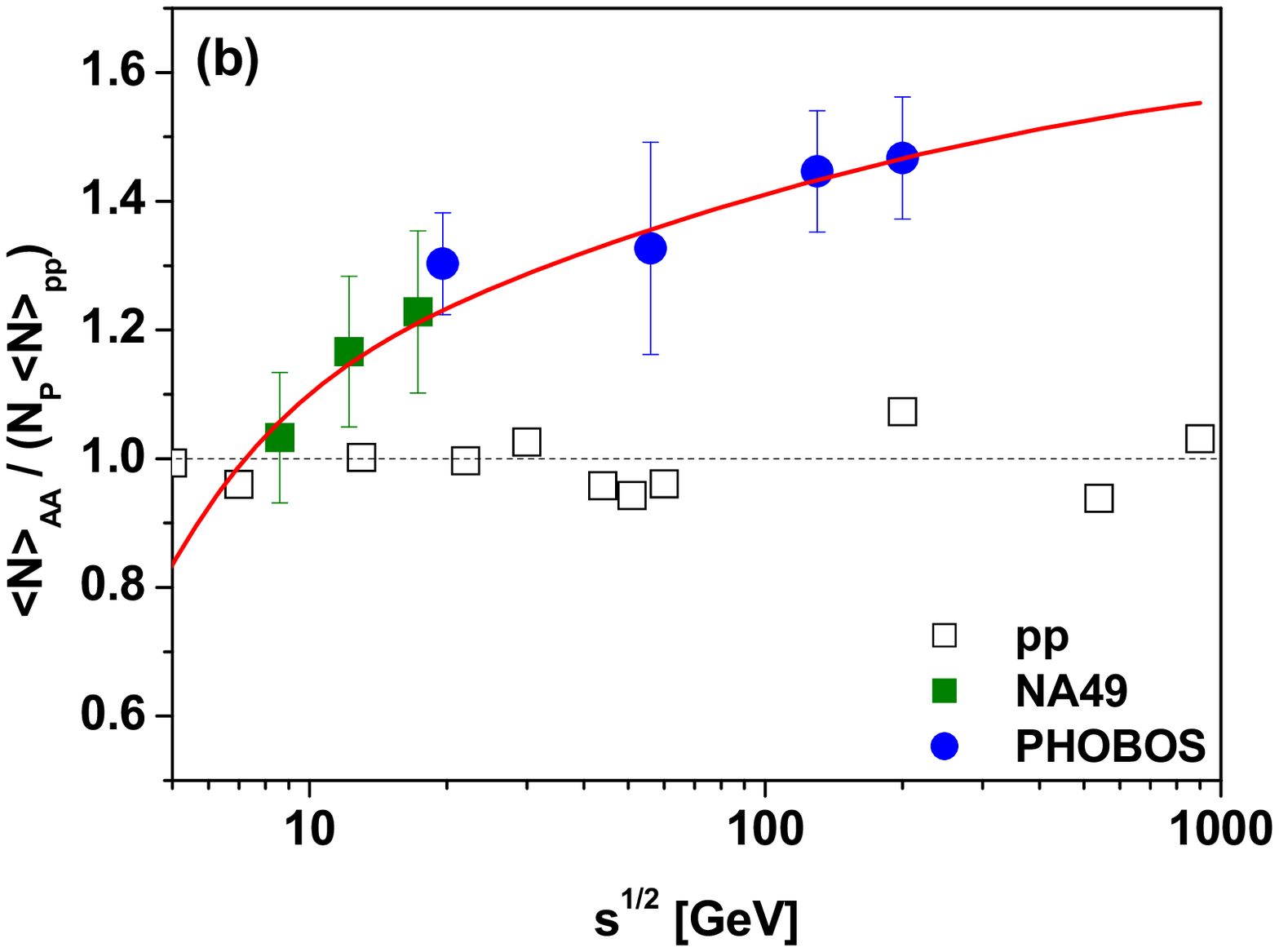}
\end{center}
\caption{(Color online) $(a)$ Energy dependencies of the
parameters $q$ obtained from, respectively: multiplicity
distributions $P(N)$ \cite{mult} (squares), from different
analysis of transverse momenta distributions $f\left( p_T \right)$
in $p+p$ data (\cite{Wibig} - circles, full symbols) and from data
on $f\left( p_T \right)$ from Pb+Pb collisions (\cite{AA} - half
filled circles). $(b)$ Energy dependence of the charged
multiplicity for nucleus-nucleus collisions divided by the
superposition of multiplicities from proton-proton collisions
fitted to data on multiplicity taken from \cite{AA} (NA49) and
from compilation \cite{PHOBOS}.} \label{Duality}
\end{figure}
\begin{equation}
S^{(\nu)}_q = \sum_{k=1}^{\nu}\frac{\nu !}{(\nu -
k)!k!}(1-q)^{k-1}\left[ S^{(1)}_q\right]^k = \frac{\left[ 1 +
(1-q)S^{(1)}_q\right]^{\nu} - 1}{1 - q}. \label{eq:Sqnu}
\end{equation}
Notice that $\ln \left[ 1 + (1 - q) S^{(\nu)}_q\right] = \nu \ln
\left[ 1 + (1 - q) S^{(1)}_q\right]$ and $S^{(\nu)}_q \stackrel{ q
\rightarrow 1}{\longrightarrow} \nu \cdot S^{(1)}_1$. For $q < 1$
entropy $S^{(\nu)}_q$ is nonextensive because $S^{(\nu)}_q/\nu
\stackrel{\nu \rightarrow \infty}{\longrightarrow} \infty$. For $
q  > 1$ one has $S^{(\nu)}_q \ge 0$ only for $q < 1 + 1/S^{(1)}_q$
and $S^{(\nu)}_q/\nu \stackrel{\nu \rightarrow
\infty}{\longrightarrow} 0$, i.e., entropy is extensive, $0 \leq
S^{(\nu)}_q/\nu \le S^{(1)}_q$.

Because $\langle N_{AA}\rangle > N_P \langle N_{pp}\rangle = \nu
 \langle N_{pp}\rangle$, the nonextensivity parameter obtained from the corresponding
entropies must be smaller than unity, $q_2 < 1$. On the other
hand, all estimates of nonextensivity parameter from Tsallis
distributions lead to $q_1 > 1$.

\section{Dressed Tsallis distributions (log-periodic oscillations)}
\label{sec:logosc}

\begin{figure} [h]
\begin{center}
\includegraphics[scale=0.35]{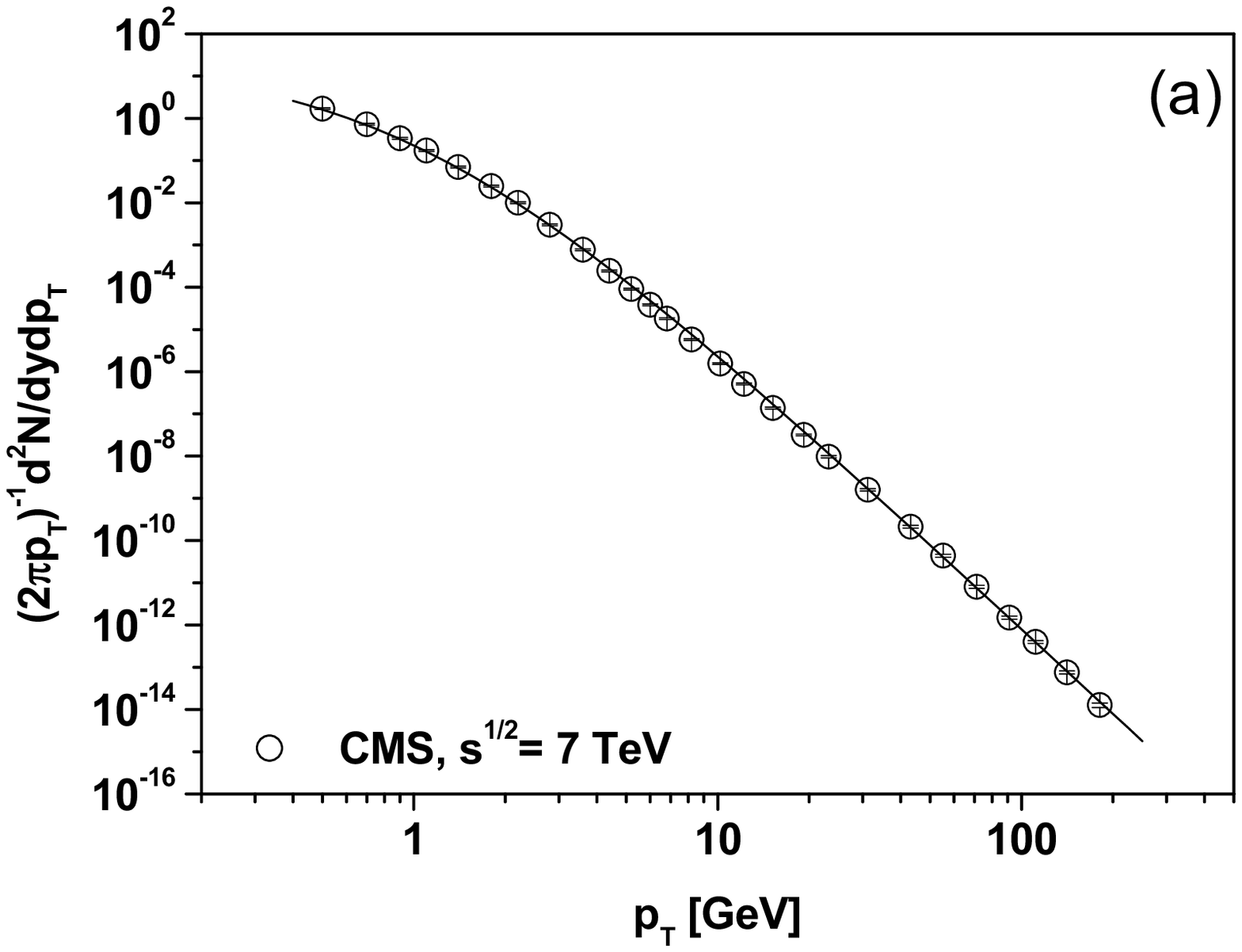}
\includegraphics[scale=0.35]{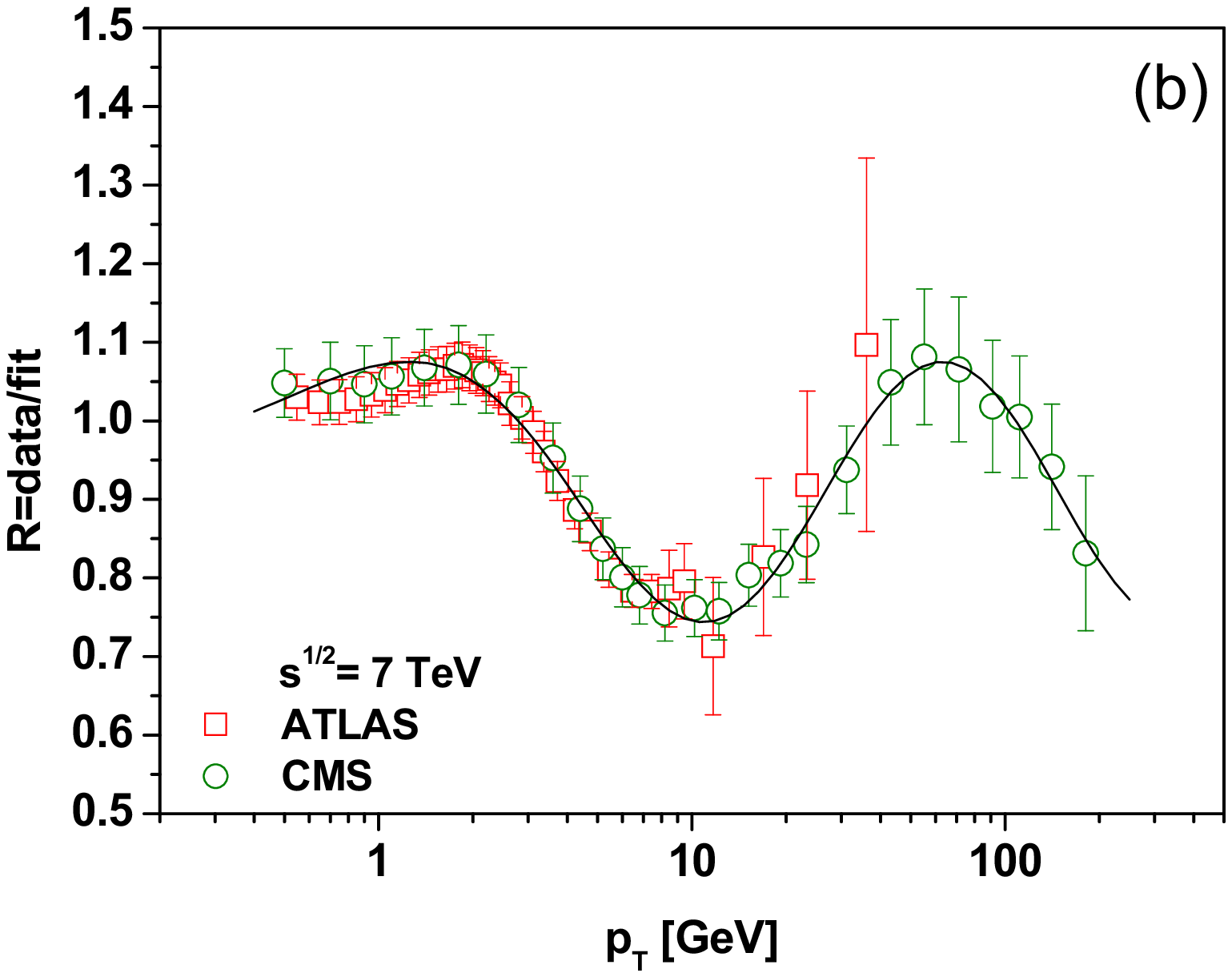}
\includegraphics[scale=0.35]{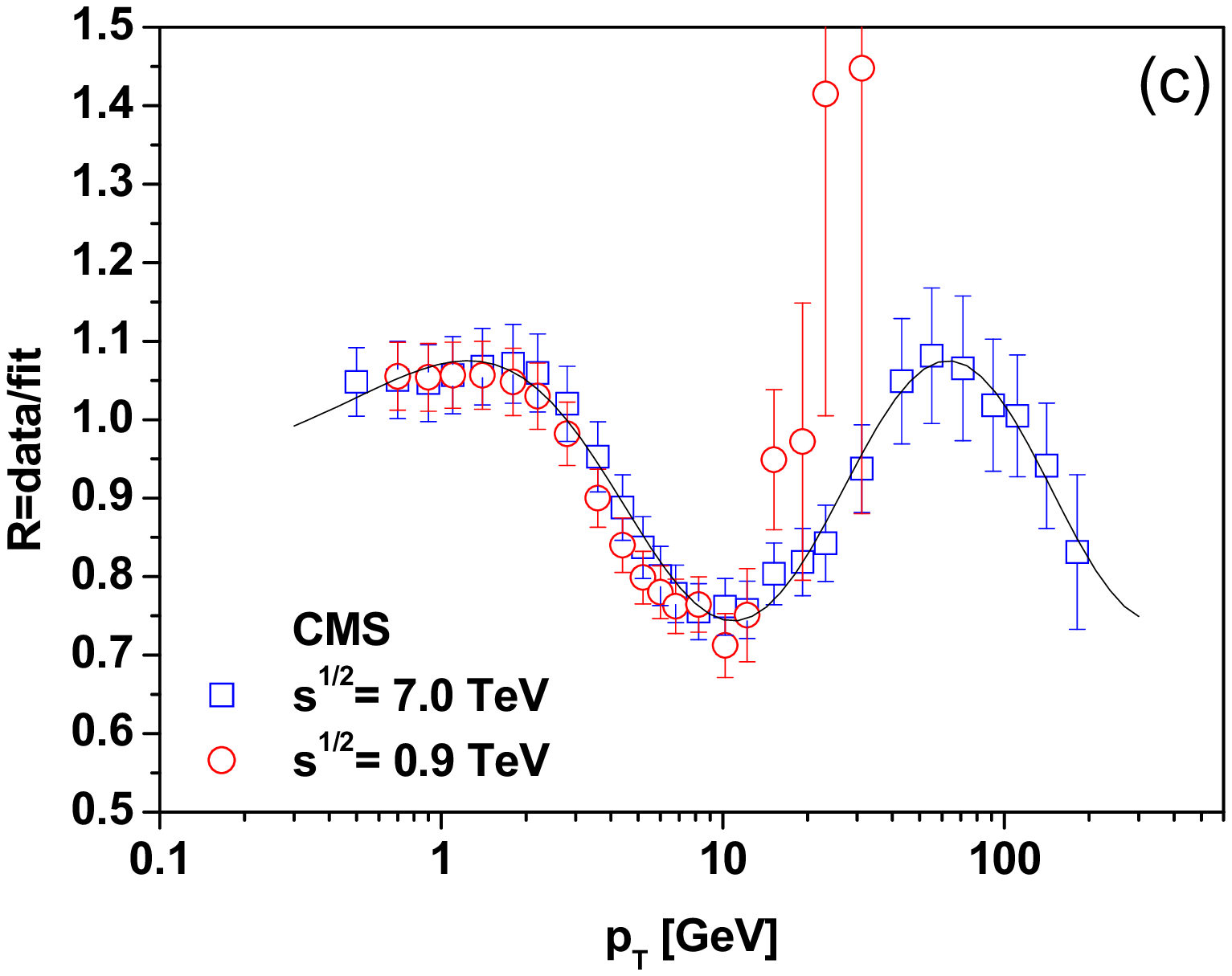}
\includegraphics[scale=0.35]{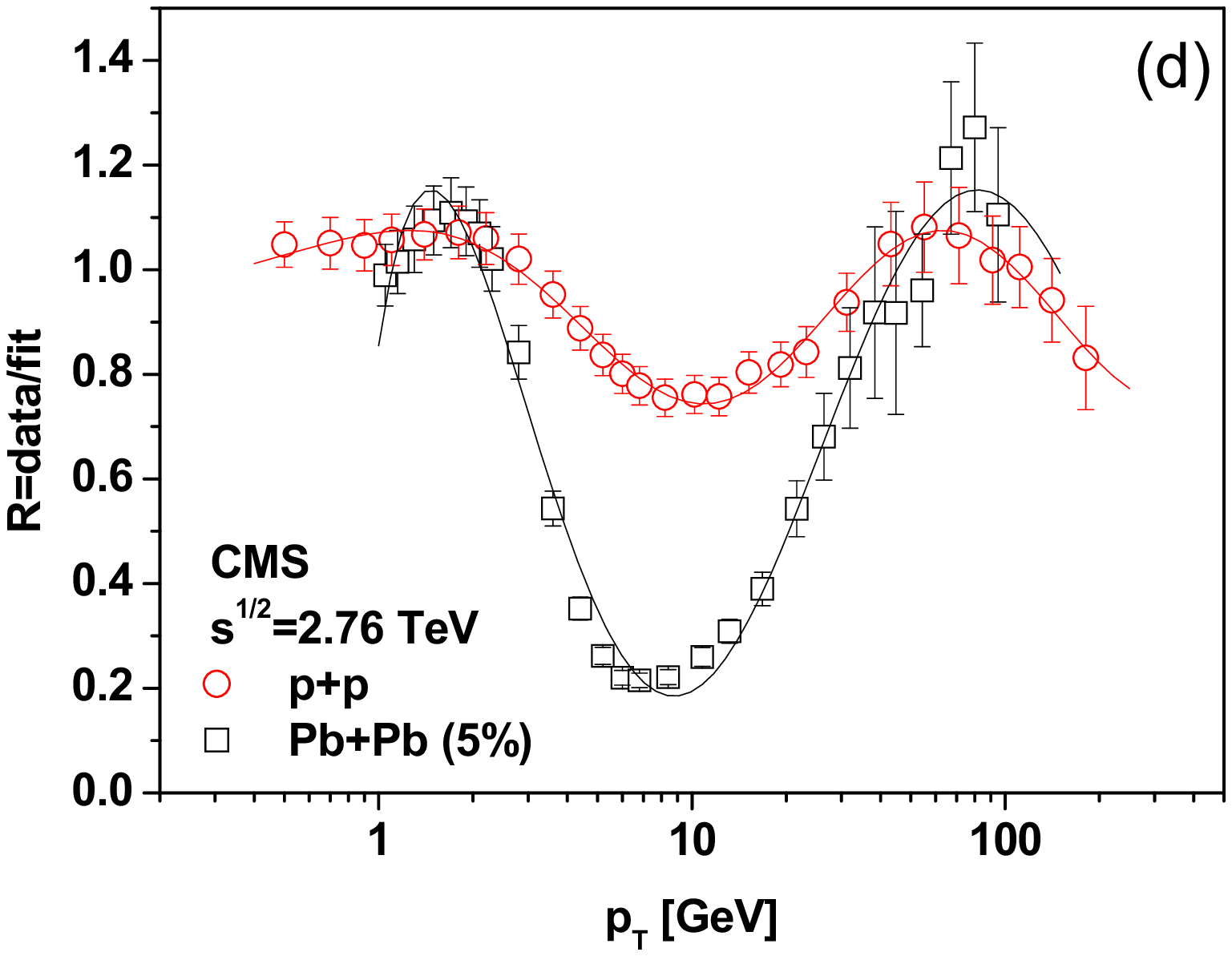}
\end{center}
\caption{ (Color online) Examples of log-periodic oscillations.
$(a)$ $dN/dp_T$ for the highest energy $7$ TeV, the Tsallis
behavior is evident. Only CMS data are shown \cite{CMS}, others
behave essentially in the identical manner. $(b)$ Log-periodic
oscillations showing up in different experimental data like
\cite{CMS} or ATLAS \cite{ATLAS} taken at $7$ TeV. $(c)$ Results
from CMS \cite{CMS} for different energies. $(d)$ Results for
different systems ($p+p$ collisions compared with $Pb+Pb$ taken
for $5$ \% centrality \cite{CMSPb}. Results from ALICE are very
similar \cite{ALICEPb}.} \label{Summary}
\end{figure}

The pure power-like distributions are known to be in many cases
decorated by specific log-periodic oscillations (i.e, multiplied
by some dressing factor $R$) \cite{LPO-examples}. They suggest
some hierarchical fine-structure existing in the system under
consideration and are usually regarded as possibly indicating some
kind of multifractality in the system. Closer inspection of recent
data from LHC \cite{CMS,ATLAS,ALICE,CMSPb,ALICEPb} reveals that,
for large transverse momenta $p_T$, one observes a similar effect,
cf. Fig. \ref{Summary}. So far, the prevailing opinion is that
this is just an apparatus induced artifact with no meaning.
However, its persistence in the type of experiment considered,
energy or type of collision process (provided that the range of
$p_T$ covered is large enough) calls for some explanation. Such an
explanation was considered in \cite{WW_LPO,WW_LPOT}. In
\cite{LPO-examples}, the only possibility investigated was to
attribute such oscillations to complex values of the power index
$n$ in Eq. (\ref{eq:H}). As shown in \cite{WW_LPO} and also below,
it also works in the case of quasi-power like Tsallis
distributions. However, because one now also has a scale parameter
$T$ and a constant term, one can also offer another explanation:
real $n$ but log-periodically oscillating scale parameter $T$.
This was discussed in detail in \cite{WW_LPOT}. We shall present
both possibilities here.

\subsection{Complex nonextensivity parameter $q$}
\label{sec:cq}

For simple power laws, one has some function $O(x)$ which is scale
invariant, $O(\lambda x) = \mu O(x)$ and $O(x) = Cx^{-m}$ with $m
= - \ln \mu/\ln \lambda$. However, this can be written as $\mu
\lambda^{m} = 1 = e^{i2\pi k}$, where $k$ is an arbitrary integer.
We then have not a single power $m$ but rather a whole family of
complex powers, $m_k$, with $m_k = - \ln \mu/\ln \lambda + i 2\pi
k/\ln \lambda$. Their imaginary part signals a hierarchy of scales
leading to log-periodic oscillations. This means that, in fact,
$O(x) = \sum_{k=0}w_k Re\left(x^{-m_k}\right) = x^{-Re \left( m_k
\right)} \sum_{k=0} w_k\cos\left[Im\left(m_k\right)\ln(x)\right]$
(where $w_k$ are coefficients of the expansion). This the origin
of the usual dressing factor appearing in \cite{LPO-examples} and
used to describe data:
\begin{equation}
R(E)= a + b\cos\left[ c\ln(E + d) + f\right] \label{eq:R}
\end{equation}
(only $w_1$ and $w_2$ terms are kept).

It turns out that a similar scaling solution can also be obtained
in case of a Tsallis quasi-power like distribution. To this end
one must start from stochastic network approach, Section
\ref{sec:Sn} and Eq. (\ref{eq:nets}), in which Tsallis
distribution is obtained by introducing a scale parameter
depending on the variable considered. In our case it is $df(E)/dE
= - f(E)/T(E)$ resulting in
\begin{equation}
f(E) = \frac{n-1}{nT_0}\left(1 +
\frac{E}{nT_0}\right)^{-n}\quad{\rm for}\quad T(E) = T_o +
\frac{E}{n.} \label{eq:rgR}
\end{equation}
In final difference form (with change in notation: $T_0$ replaced
by $T$)
\begin{equation}
\frac{df(E)}{dE} = - \frac{f(E)}{T(E)}~~ \Longrightarrow ~~f(E +
\delta E) = \frac{ - n\delta E + nT +E}{nT + E} f(E).
\label{eq:deltaE}
\end{equation}
We consider a situation in which $\delta E = \alpha nT(E) = \alpha
(nT + E)$. It depends now on the new scale parameter $\alpha$
($\alpha < 1/n$ in order to keep changes of $\delta E$ to be of
the order of $T$) and can be very small but always remains finite.
It can now be shown that $f[E + \alpha(nT + E)] = (1 - \alpha
n)f(E)$ which, when expressed in the new variable $x = 1 +
E/(nT)$, corresponds formally to the following scale invariant
relation:
\begin{equation}
g[(1 + \alpha) x] = (1 - \alpha n)g(x). \label{eq:gscaling}
\end{equation}
Following the same procedure used to obtain dressed solutions of
scale invariant functions discussed at the beginning of this
Section, one arrives at the dressed Tsallis distribution (we keep,
as before, only the two lowest terms, $k=0$ and
$k=1$)\footnote{Notice that in Eq. (\ref{eq:approx}) $n \neq m_0$.
However, $n$ and $T$ are both unknown {\it a priori} parameters.
Therefore, for fitting purposes, where we have to use two (and not
three) parameter Tsallis distribution, we use $[ 1 +
E/(m_0T')]^{-m_0}$ with fitting parameters $m_0$ and $T'$. In
terms of $T$, $n$ and $\alpha$ we have $T' = nT/m_0  = -nT\ln( 1 +
\alpha)/\ln( 1 - \alpha n)$.}:
\begin{equation}
g(E) \simeq \left( 1 + \frac{E}{nT}\right)^{-m_0}\left\{ w_0 +
w_1\cos\left[ \frac{2\pi}{\ln (1 + \alpha)} \ln \left( 1 +
\frac{E}{nT}\right)\right]\right\}. \label{eq:approx}
\end{equation}
with $m_0 = - \ln(1 - \alpha n)/\ln(1 + \alpha ) \stackrel{\alpha
\rightarrow 0}{\longrightarrow} n$. In addition to the scale
parameter $\alpha$ one has two more parameters occurring in the
dressing factor $R$, $w_0$ and $w_1$. The other parameters
occurring in Eq. (\ref{eq:R}) are expressed by the original
parameters in the following way: $a/b = w_0/w_1$, $c = 2\pi/\ln(1
+ \alpha)$, $d=nT$ and $f = -2\pi\ln(nT)/\ln(1+\alpha) = -c\ln d$.
One can, however, consider a more involved evolution process, with
$\kappa$ sequential cascades; in this case the additional
parameter $\kappa$ changes parameter $c$ in (\ref{eq:R}), $c
\rightarrow c'=c/\kappa$. It does not affect the slope parameter
$m_0$ but changes the frequency of oscillations which now decrease
as $1/\kappa$. Comparison with data requires $\kappa \sim 22$ (cf.
\cite{WW_LPO} for details).

\subsection{Log-periodically oscillating $T$}
\label{sec:LPOT}

\begin{figure} [t]
\begin{center}
\includegraphics[scale=0.37]{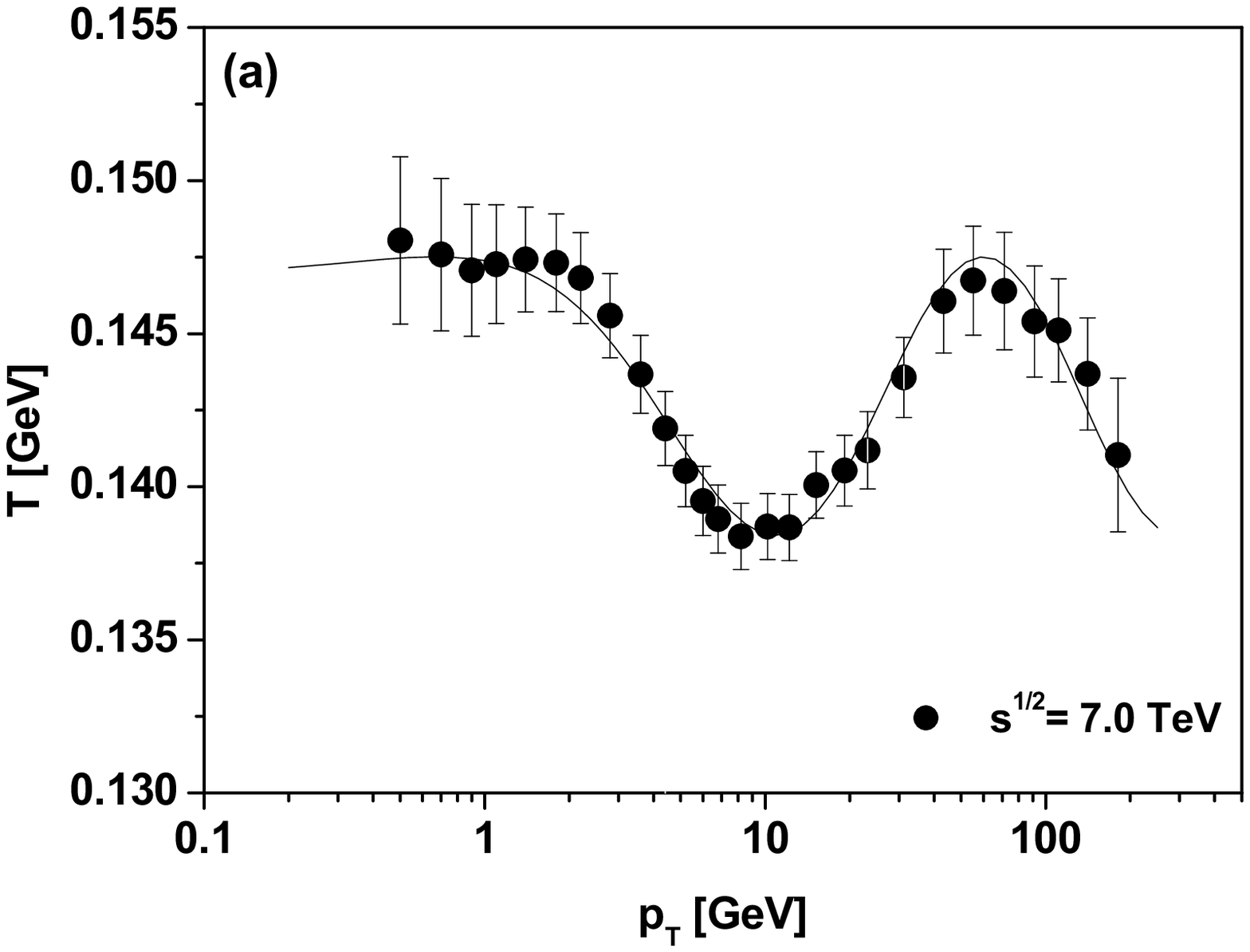}
\includegraphics[scale=0.37]{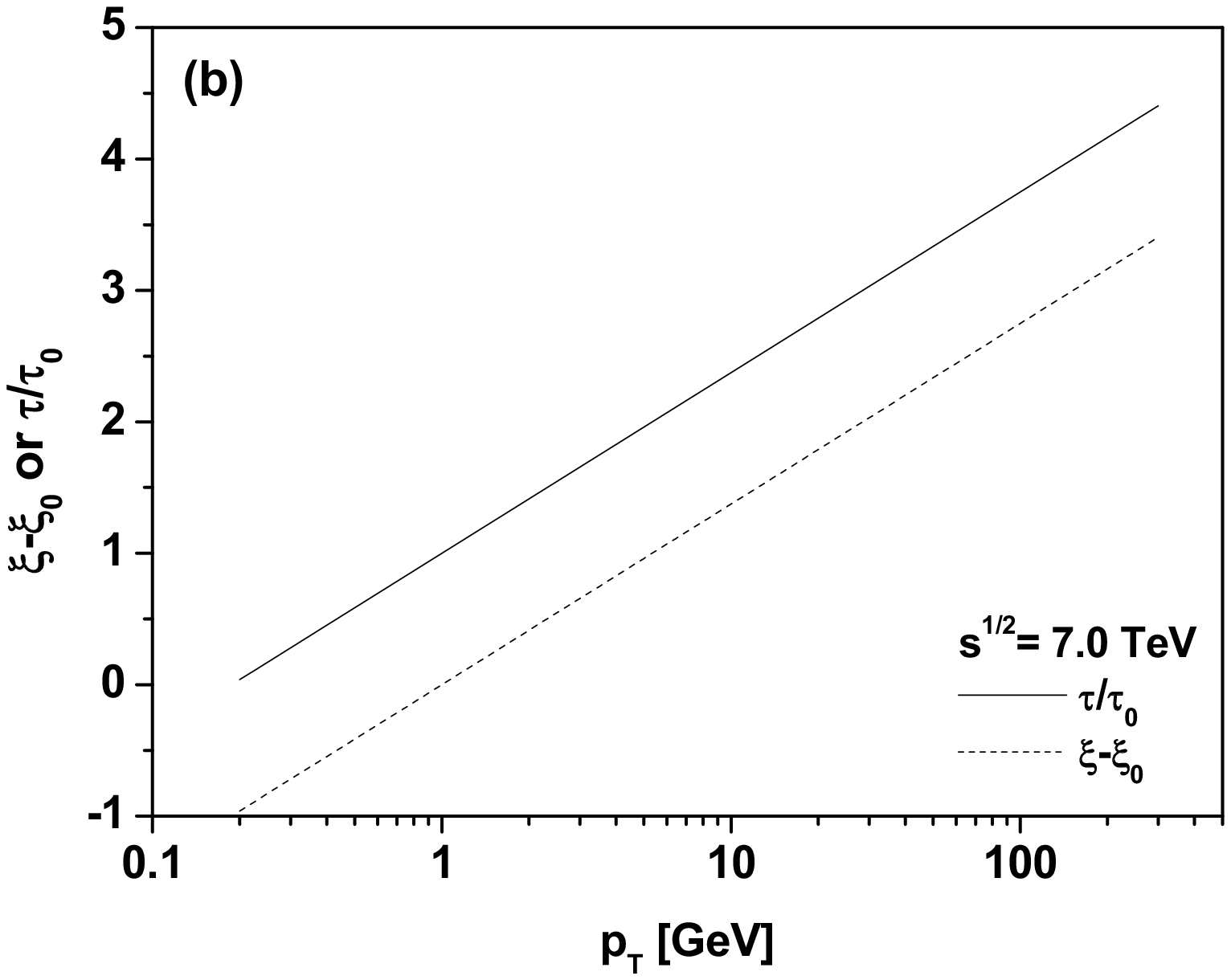}
\end{center}
\caption{ (Color online) $(a)$ Oscillations of scale parameter $T$
leading to identical dressed Tsallis distribution as shown in Fig.
\ref{Summary}b (obtained for CMS data at $7$ TeV and fitted using
Eq. (\ref{eq:T})). $(b)$ Dependencies of $\tau/\tau_0$ from Eq.
(\ref{eq:tauE}) and $\xi - \xi_0$ from Eq. (\ref{eq:St6})
resulting in oscillations of $T$ shown in panel $(a)$.
 } \label{FigLPOT}
\end{figure}

As mentioned before, one can translate a dressed Tsallis
distribution into a normal one but with a log-periodically
oscillating in $p_T$ scale factor $T$, cf. Fig. \ref{FigLPOT}a.
The formula used there to fit the obtained results resembles that
for dressing factor (\ref{eq:R}),
\begin{equation}
T = \bar{a} + \bar{b}\sin\left[ \bar{c} \left( \ln(E +
\bar{d}\right) + \bar{f} \right]. \label{eq:T}
\end{equation}
In fit shown in Fig. \ref{FigLPOT}a parameters (generally energy
dependent) are $\bar{a} = 0.143,~\bar{b} = 0.0045,~\bar{c} =
2.0,~\bar{d} = 2.0,~\bar{f} = -0.4$.

To explain Eq. (\ref{eq:T}) one uses a stochastic equation for the
temperature evolution \cite{Kampen} written in Langevin
formulation with energy dependent noise, $\xi(t,E)$, and allowing
for time dependent $E=E(t)$ ~~\footnote{Notice the change of
notation, we discuss formulas for energy $E$ but results are for
transverse momenta $p_T$ here. However, they are taken at the
midrapidity, i.e., for $y \simeq 0$, and for large transverse
momenta, $p_T > M $, and in this region one has $E=\sqrt{M^2 +
p^2_T} \cosh(y) \simeq p_T$.}:
\begin{equation}
\frac{dT}{dE}\frac{dE}{dt} + \frac{1}{\tau} T + \xi(t,E) T = \Phi.
\label{eq:St2}
\end{equation}
Assuming now a scenario of {\it preferential attachment} (cf.
Section \ref{sec:Sn} above) known from the growth of networks
\cite{nets}) one has
\begin{equation}
\frac{dE}{dt} = \frac{E}{n} + T. \label{eq:St3}
\end{equation}
and Eq. (\ref{eq:St2}) has now the form:
\begin{equation}
\left(\frac{E}{n} + T\right)\frac{dT}{dE} + \frac{1}{\tau} T +
\xi(t,E) T = \Phi .\label{eq:St4}
\end{equation}
After straightforward manipulations (cf. \cite{WW_LPO} for
details) one gets, for large $E$ (i.e., neglecting terms $\propto
1/E$):
\begin{equation}
\frac{1}{n}\frac{d^2T}{d(\ln E)^2} + \left[ \frac{1}{\tau} +
\xi(t,E) \right]\frac{dT}{d(\ln E)} + T\frac{d\xi(t,E)}{d(\ln E)}
= 0.\label{eq:St5}
\end{equation}
Assuming now that noise $\xi(t,E)$ increases logarithmically with
energy,
\begin{equation}
\xi(t,E) = \xi_0(t) + \frac{\omega^2}{n}\ln E. \label{eq:St6}
\end{equation}
In this case Eq. (\ref{eq:St5}) becomes an equation for the damped
hadronic oscillator with solution in the form of log-periodic
oscillation of temperature with frequency $\omega$ and depending
on initial conditions phase shift parameter $\phi$:
\begin{equation}
T = C \exp\left\{ - n\cdot\left[\frac{1}{2\tau} +
\frac{\xi(t,E)}{2}\right]\ln E\right\}\cdot \sin(\omega\ln E +
\phi). \label{eq:St7}
\end{equation}
Averaging the noise fluctuations over time $t$ and taking into
account that the  noise term cannot on average change the
temperature, $1/\tau + \langle \xi(t,E)\rangle = 0$, one arrives
at
\begin{equation}
T = \bar{a} + \frac{b'}{n}\sin( \omega \ln E + \phi).
\label{eq:St9}
\end{equation}
This should now be compared with the parametrization of $T(E)$
given by Eq. (\ref{eq:T}) and used to fit data in
Fig.\ref{FigLPOT}~~\footnote{Notice that only a small amount of
$T$, of the order of $\bar{b}/\bar{a} \sim 3\%$, emerges from the
stochastic process with energy dependent noise; the main
contribution comes from the usual energy-independent Gaussian
white noise.}.

We close with the remark that, instead of using energy dependent
noise $\xi(t,E)$ given by  Eq. (\ref{eq:St6}) and keeping the
relaxation time $\tau$ constant. we could equivalently keep the
energy independent white noise, $\xi(t,E) = \xi_0(t)$, but allow
for the energy dependent relaxation time, for example in the form
of
\begin{equation}
\tau = \tau(E)  = \frac{n\tau_0}{n + \omega^2 \ln E}.
\label{eq:tauE}
\end{equation}
In this case the temperature evolution has the form
\begin{equation}
T(t) = \langle T\rangle + [T(t=0) - \langle T\rangle]
E^{-t\omega^2/n} \exp\left(-\frac{t}{\tau_0}\right),
\label{eq:Ttau}
\end{equation}
and $T$ gradually approaches its equilibrium value $\langle T
\rangle$. Actually, for $\tau = \tau(E)$, as in our case, this
approach towards equilibrium is faster for large $E$. This is
because, in addition to the usual exponential relaxation
characteristic for $\tau = const$ case, we have an additional
factor $\sim E^{- t \omega^2/n}$.

\section{Summary and conclusions}
\label{sec:Summary}

We presented examples of possible mechanisms resulting in
quasi-power distributions exemplified by Tsallis distribution, Eq.
(\ref{eq:Tsallis}). Our presentation had to be limited, therefore
we did not touch thermodynamic connections of this distribution
\cite{Sq,consistency} or the possible connection of Tsallis
distributions with QCD calculations discussed recently
\cite{qQCD,qWWCT}.

The main results presented here can be summarized in the following
points:
\begin{itemize}
\item Statistical physics consideration, as well as "induced
partition process", results in Eq. (\ref{eq:ConstN}), i.e., in
Tsallis distribution with $q = (N - 3)/(N - 2) < 1$. Fluctuations
of the multiplicity $N$ modify the parameter $q$ which is now
equal to $q = 1 + Var(N)/\langle N\rangle^2 - 1/\langle N\rangle$,
cf. Eq. (\ref{eq:FluctN}). Notice that conditional probability for
the BG distribution again results in Eq. (\ref{eq:ConstN}).

\item Fluctuations of the multiplicity $N$ are equivalent to
results of an application of superstatistics, where the
convolution
\begin{equation}
f(E) = \int g(T) \exp\left( - \frac{E}{T}\right) dT \label{eq:SSt}
\end{equation}
becomes a Tsallis distribution, Eq. (\ref{eq:Tsallis}), for
\begin{equation}
g(T) = \frac{1}{\Gamma(n)T}\left(\frac{nT_0}{T}\right)^n
\exp\left( - \frac{nT_0}{T}\right). \label{eq:gamma}
\end{equation}
\item Differentiating Eq. (\ref{eq:SSt}) one gets
\begin{equation}
\frac{df(E)}{dE} = - \frac{1}{T(E)} f(E)\qquad {\rm where}\quad
T(E) = T_0 + \frac{E}{n}. \label{eq:diffSSt}
\end{equation}
This is nothing else than a "preferential attachment" case, again
resulting in a Tsallis distribution which for $T(E) = T_0$ becomes
a BG distribution, cf. Eq. (\ref{eq:nets}).

\item Replacing  in Eq. (\ref{eq:diffSSt}) differentials by finite
differences, cf. Eq. (\ref{eq:deltaE}), one gets for $\delta E =
\alpha n T(E)$ the scale invariant relation, Eq.
(\ref{eq:gscaling}), which results in log-periodic oscillations in
Tsallis distributions \footnote{Among numerous other explanations
we can therefore say that we have demonstrated that a Tsallis
distribution, which can be regarded as generalization to real
power $n$ of such well known distributions as the Snedecor
distribution (with $n = (\nu + 2)/2 $ with integer $\nu$, for $\nu
\rightarrow \infty$ it becomes an exponential distribution), can
be extended to complex nonextensivity parameter.}.

\end{itemize}

In addition to this line of reasoning, we have also brought in the
problem of the apparent duality between the nonextensive
parameters obtained from the whole phase space measurements of
multiplicity and more local measurements of transverse momenta.
This point deserves an experimental and phenomenological scrutiny.

Finally, we tentatively suggested that, by choosing the right
constraints, which account for additive or multiplicative
processes considered, one can also get a Tsallis distribution
directly from the Shannon information entropy.

\vspace*{0.3cm} \centerline{\bf Acknowledgment}

\vspace*{0.3cm} This research  was supported in part by the
National Science Center (NCN) under contract
DEC-2013/09/B/ST2/02897. We would like to warmly thank Dr Eryk
Infeld for reading this manuscript.

\end{document}